\documentclass[twocolumn,aps,prb,superscriptaddress]{revtex4-2}

\usepackage{graphicx}
\usepackage{amssymb}

\begin{document}

\preprint{preprint}

\title{II. Exploring the role of the Crystal Electric Field in the vicinity of a\\ Quantum Critical Point}



\author{J.G. Sereni*}
\affiliation{Low Temperature Division, CAB-CNEA, CONICET, IB-UNCuyo, 8400 Bariloche, Argentina}

\author{I. Čurlik}
\affiliation{Faculty of Sciences, University of Prešov, 17. novembra 1, SK - 080 78 Prešov, Slovakia}

\author{S. Gabani}
\affiliation{Institute of Experimental Physics, SAS, Watsonova 47, SK - 040 01 Košice, Slovakia}

\author{M. Giovannini}
\affiliation{Department of Chemistry, University of Genova, Via Dodecaneso 31, Genova, Italy}


\email[]  {jsereni@yahoo.com}

\date{\today}

\begin{abstract}

Very low temperature thermodynamic properties of the Yb$M_{5-x}X_x$ (with $M$= Ni, Cu, and $X$= Cd, Mg, Au, Zn, Ag) cubic compounds are analyzed covering a broad range of behavior  
between magnetic and Fermi-liquid ground states GS using the {\it chemical doping}: $\zeta$ as control parameter. This allows to gain insight into the evolution of the GS behavior including a quantum critical point QCP. 
Doniach-Lavagna phase diagram limitations are improved by taking into account crystal electric field CEF splittings.  
Three regions are recognized as a function of $\zeta$: i) a magnetic one with long range magnetic order and $T_{ord}\propto T_N \approx 1$\,K, that weakens the interactions between $0.8\,K  
\geq T_m\geq 0.4$\,K, exhibiting very low Kondo temperature $T_K^{GS}$ in respective doublets GS. Then, for $T_Q\leq 0.4$\,K, quantum fluctuations start to dominate the scenario with the 
specific heat  $C_{4f}/T(T\geq T_Q)$ showing $T$ power law dependencies, and a very heavy-fermion {\it plateau} below $T_Q$. ii) beyond the QCP the typical Non-Fermi-Liquid 
logarithmic $T$ dependence: $C_{4f}/T \propto \ln(T/T_0)$, with traces of magnetic order. At the non-magnetic limit: iii) the alloys behave as valence-fluctuation systems with growing $T_K$ that 
overcomes the CEF splitting. 
With this experimental information, a realistic phase diagram can be drawn around the QCP where the scenario is dominated by low lying quantum fluctuations, without $C_{4f}/T|_{Lim T\to 0}$ 
divergences but a clear drop entering into the non-magnetic phase. 

\end{abstract}


\maketitle


\begin{figure}
\begin{center}
\includegraphics[width=18pc]{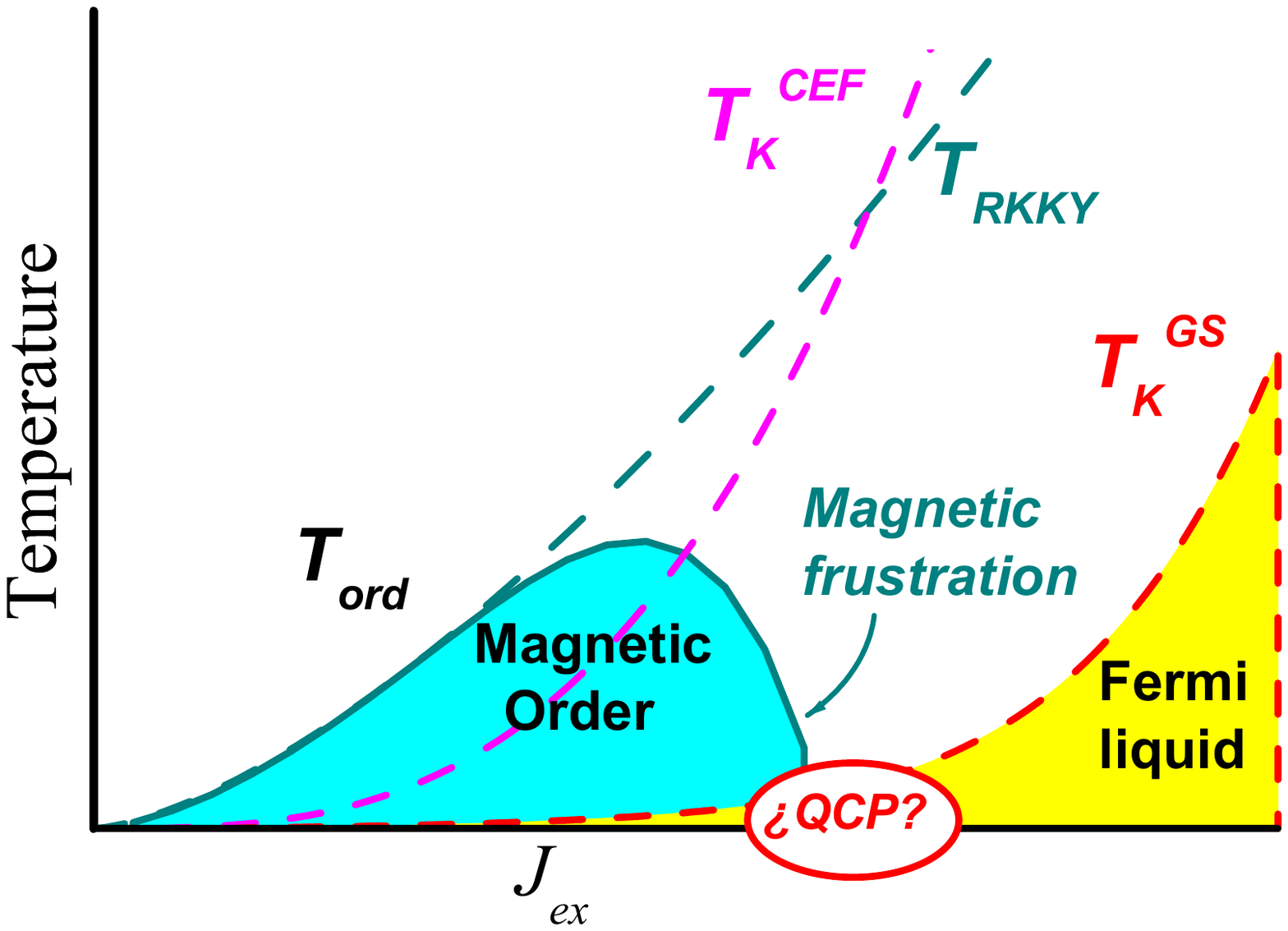}
\caption{Doniach-Lavagna phase diagram as a function of the magnetic exchange, showing the different variation between the full Kondo temperature: $T_K^{CEF}(J_{ex})$, used in that model, and the actual ground state: $T_K^{GS}$. The magnetic phase  boundary $T_{ord} \propto T_{RKKY} \propto J_{ex}$  \cite{RKKY}.}

\label{F1}
\end{center} 
\end{figure}

\section{Introduction}

The earliest approach to describe the transition between magnetic and non-magnetic phases dominated by Kondo screening, was proposed by Doniach 
\cite{Doniach} and Lavagna \cite{Lavagna} D-L about four decades ago. Within that model, the dependence on a unique magnetic exchange parameter $J_{ex}$ allowed to compare the intensity 
of the ordering magnetic interaction: $T_{ord} \propto T_{RKKY} \propto J_{ex}^2$  \cite{RKKY} and the Kondo effect: $T_K\propto \exp(-1/ J_{ex})$, as shown in the schematic phase diagram in 
Fig.~\ref{F1}. 

In this diagram there are two possible $J_{ex}$ values for which $T_{ord} \to 0$, the trivial limit of $J_{ex}\to 0$ and where the Kondo effect fully screens the magnetic moment driving $
\mu_{eff}\to 0$.
The latter is currently identified as a quantum critical point QCP because, in theory, it was defined as the point where a phase boundary extrapolates to $T=0$. 
\cite{QCP,Si1}. However, 
such hypothesis dodges the fact that screened $\mu_{eff}\to 0$ makes the transition to vanish suppressing the phase boundary itself. This occurs even at $T\geq 0$ \cite{PhilMag} in several 
pressure induced antiferromagnetic superconductors \cite{CePd2Si2}. 
An improvement to this model was done by recognizing the difference between the magnetic exchange $J_{RKKY}$  mediated by conduction electrons and the {\it 4f-band electrons} 
interaction $J_K$ that hybridizes those 
states \cite{Iglesias}. On the contrary, the presence of robust moments is required to properly define a phase boundary down to $T\to 0$ t the QCP.

A more realistic approach can be done by taking into account the crystal electric field CEF splitting on the $4f$ levels ($\Gamma_6, \Gamma_7$ and $\Gamma_8$ for $J=7/2$ -Yb$^{3+}$ ions). 
That approach is based on the  
fact that each $\Gamma_i$ state may have a different Kondo {\it 4f-band} hybridization factor. 
Despite for simplicity such a difference is usually not taken into account, the fact that different Hund's-rule split levels are irreducible representations shall have different hybridization factors.  

Such splitting provides the possibility to find a level scheme where the double-GS keeps well defined magnetic moments, i.e. with low $T_K^{GS}$ values, while the excited levels may develop  
significant hybridization with relevant $T_K^{CEF}$ values.  
 In Fig.~\ref{F1} we represent two trajectories for $T_K$, with the $T_K^{CEF}$ growing as proposed in the standard D-L  model, and $T_K^{GS}$ which starts to grow at larger values of 
 $J_K$. Along with that condition, it 
 is necessary to have weak magnetic interactions to allow to suppress the magnetic order as $T_{ord}\to 0$ or, eventually, some topological conditions like geometric or magnetic frustration \cite{geofrustr} that inhibit long 
 range interactions development. In the studied Yb$M_4X$ compounds, the weakness of the $J_{RKKY}$ interaction arises from the significant magnetic Yb-Yb distance and the fact that the 
 interaction is mediated by Yb-$M$ and Yb-$X$ shared band electrons of first and second Yb atomic neighbors \cite{Poettgen}.  

On the high $J_{ex}$ values side of the D-L phase diagram, the exponential increase of $T_K(J_{ex})$ \cite{Doniach, Lavagna} induces a $T_K^{CEF}$ with $T_K^{GS}$ overlap once they overcome the CEF splitting 
at the onset of the {\it Valence Fluctuation} VF regime \cite{Systematic}. 
Therefore, only a {\it 4f} rare earth like Yb that may show localized as well as incipient {\it 4f-band} states hybridization can fulfill these conditions when the energy of the CEF levels splitting exceeds 
respective $T_K$ values, i.e. $\Delta_{CEF}>T_K$ .

\begin{figure}
\begin{center}
\includegraphics[width=18pc]{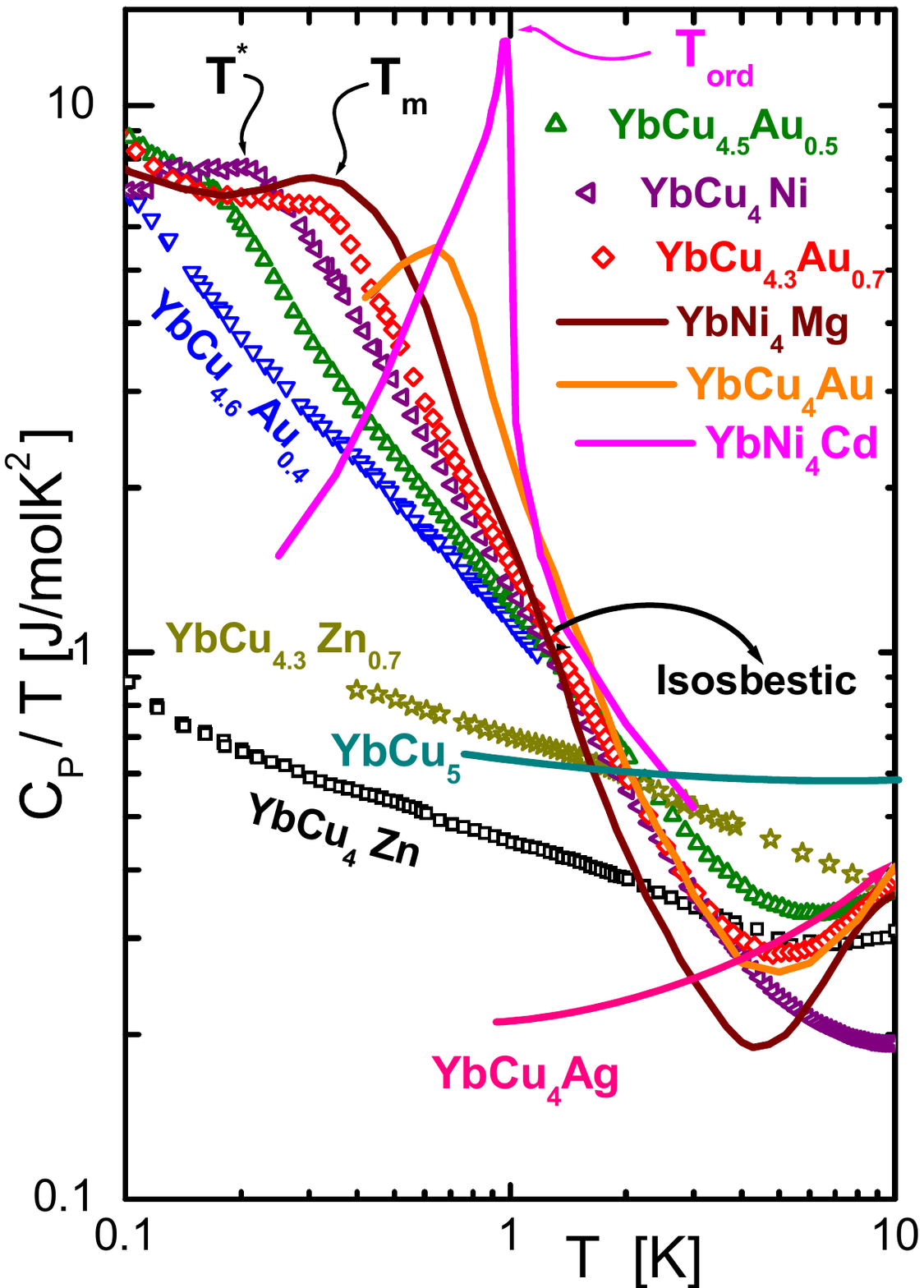}
\caption{(Color online) Comparison of the low temperature specific heat divided temperature of eight Yb$M_4X$ compounds and alloys ($M$ = Ni, Cu and $X$ = Mg \cite{YbNi4Mg}, Ni 
\cite{YbCu4Ni}, Cu \cite{YbCu5}, Au \cite{YbCu5-xAux}, Zn \cite{ArXivZn})  lying in the vicinity of a QCP, in a double logarithmic representation. 
Other compounds and alloys are not included for clarity because their $C_m/T(T)$ dependencies  overlap some of those included in the figure. 
Note that an isosbestic point \cite{isosb} occurs at $C_m/T(1.2\,K) \approx 1$\,J/molK$^2$ where respective entropy derivatives coincide. \label{F2}}
\end{center}
\end{figure}

\section{Analysis of the GS of Yb-systems in the proximity of a QCP}

Strictly, for a comprehensive analysis of the D-L phase diagram in light of experimental information, a system comprising the full range of behavior from mean field type magnetic order to non-
magnetic VF including a QCP is required. Such a broad range is hardly to be covered by usual control parameters like pressure and even alloying. 
Among the large amount of Yb-based intermetallic compounds only a few of them fulfill such conditions like the low $T_K^{GS}$ values with the magnetic order vanishing driven by chemical doping.
Particularly the family of cubic compounds: Yb$M_4X$, with $M$ = Ni, Cu and $X$= Ni  \cite{YbCu4Ni}, Au \cite{GiovLattParAu,YbCu5-xAux}), Zn \cite{Serrao}, Mg 
\cite{YbNi4MgLatPar,YbNi4Mg} and Cd \cite{YbNi4Cd,Pd}, together with cubic-YbCu$_5$\cite{YbCu5,Hamdaoui} and YbCu$_{4.5}$  \cite{Lilian4.5} (both synthetized under pressure to obtain the cubic structure) 
present these characteristics and are studied in this 
work. 

Along with the mentioned stoichiometric compounds, three series of alloys were studied providing nearly continuous information about the evolution of the magnetic properties on both sides of the QCP. 
On the magnetic one it is 
YbCu$_{5-x}$Au$_x$ \cite{YbCu4AuBau,YbCu5-xAux}, within the {\it Non-Fermi-liquid} NFL region YbCu$_{5-x}$Zn$_x$ \cite{ArXivZn}, and YbCu$_{5-x}$Ag$_x$ \cite{YbCuAgPRB,YbCuAgJalco,Hamdaoui} on 
the Fermi-liquid FL limit.  
In the following Section we will analyze respective specific heat behaviors in order to identify their each GS character. 

In all cases, the topological conditions for magnetic frustration \cite{Collem} is provided by the fcc structure of MgCu$_4$Sn type (ternary derivative of the cF24-AuBe$_5$ type structure) that can be viewed as a network of 
edge-sharing tetrahedra with Yb magnetic ions located at the vertices, like a 3D analog of triangular lattice \cite{XtlChem,ActPol}. 

\subsection{Specific heat behavior} 

\begin{figure}
\begin{center}
\includegraphics[width=20pc]{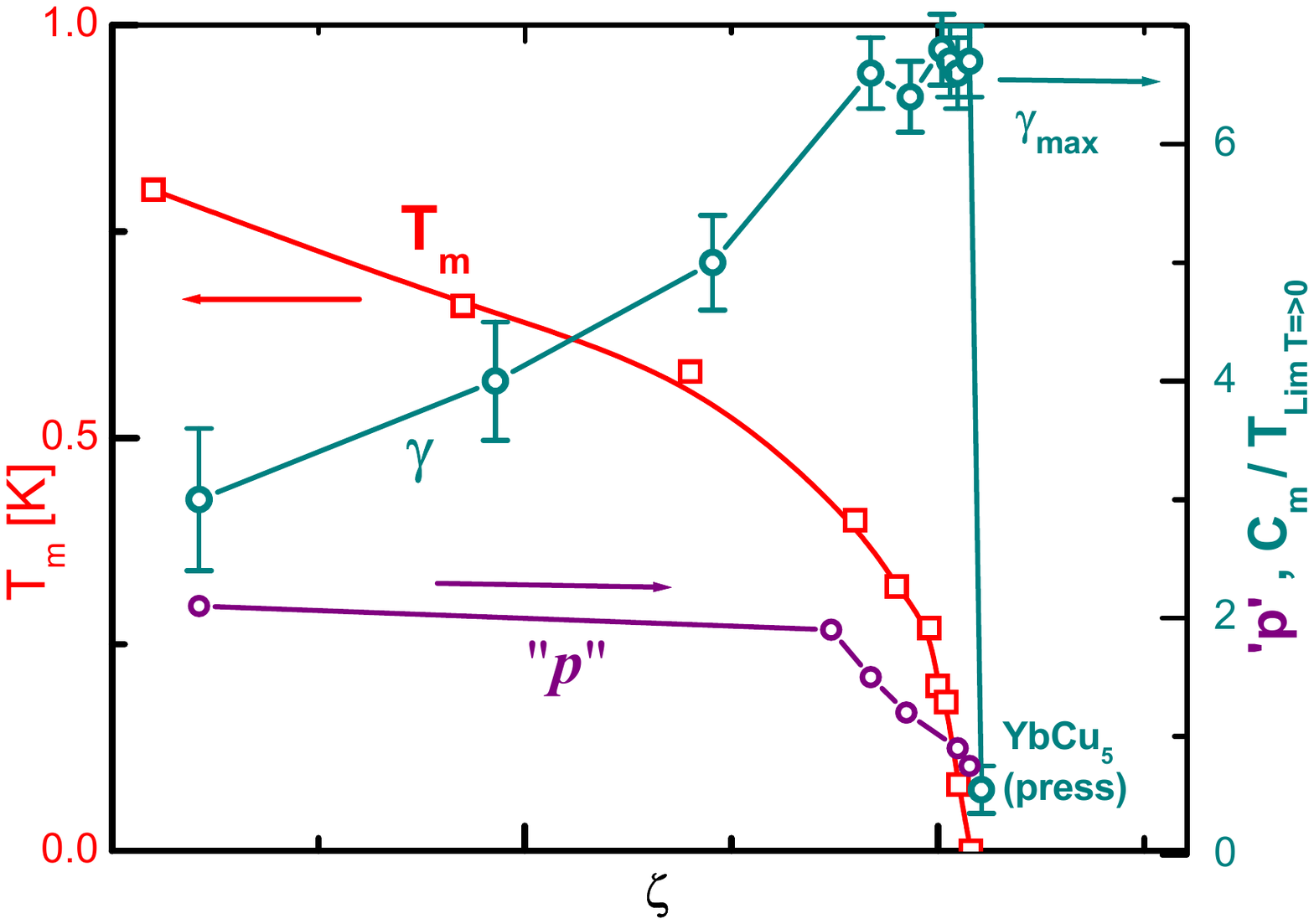}
\caption{(Color online) 
Left (red) axis: evolution of the $T_m$ anomaly (see Fig.~\ref{F1}) towards the QCP as a function of the chemical doping parameter $\zeta$ defined in the Supplementary Material. Right (cian) axis: 
increase of the $C_P/T|_{\lim{T \to 0}}=/gamma(\zeta)$ 
extrapolation till to reach a maximum $/gamma_{max}$ close to the 
QCP, and (purple) triangle: exponent '$p$' of the  $C_P/T(T>T_m)$ power laws dependence extracted from Fig 2.  \label{F3}}
\end{center}
\end{figure}

Among the mentioned Yb$M_4X$ compounds, only some of them were studied down to the $T\leq 1$\,K range. In  Fig.~\ref{F2} we compare the very low temperature specific heat of those  
compounds approaching the QCP from both sides. The 
incipient turn up below $T\approx 0.1$\,K indicates the onset  of the nuclear contribution \cite{ArXivZn}.
The comparison of the $C_P(T)/T$ dependencies between these isotypic compounds offers a unique opportunity to recognize in detail the thermodynamic behavior of real systems 
in the vicinity of a QCP. 

Different types of $C_P/T(T)$ dependencies can be clearly distinguished. The compounds with $T_N \approx 1$\,K show a transition that successively transforms into a broad 
peak at $T_m$, and then into a shoulder for $T_Q \leq 0.4$\,K. At $T > T_m$, the compounds with $M$ = Mg, Ni and Au show power law $T$ dependencies: $C_m/T \propto 1/T^p$, followed by a 
{\it plateau}: $C_P/T|_{\lim{T \to 0}}=\gamma_{max}\approx 6.7 \pm 0.4$\,J/mol K$^2$ \cite{JLTP2018}.
This feature reveals a crossover from random magnetic interactions to a nearly constant density of 
excitations regime. 

Differently, the alloys with $M$= Zn show a logarithmic $T$ dependence with lower $C_P/T|_{lim T\to 0}$  values. 
Similar scenario was observed about two decades ago in the isotypic UCu$_{5-x}$Pd$_x$, where the crossover between power law and logarithmic dependencies was already identified 
\cite{Scheidt, Maple}. Therefore, the definition for a QCP as the limit of a transition extrapolated to zero \cite{QCP} does not seem to be related to any singularity but more likely to a {\it crossover}  
between two different magnetic regimes of the GS in a context dominated by quantum fluctuations. 
Finally, a third region is represented in Fig.~\ref{F2} by YbCu$_{5-x}$Ag$_x$ and YbCu$_5$ as examples of non-magnetic or VF behavior.  
Interestingly, one can see in that figure that as $T_m \to T_Q$, the extrapolation of $C_P/T(T\to 0$:
$C_P/T|_{\lim{T \to 0}}$  tends to a $\gamma_{max}$ value. Coincidentally, once $T_m = T_Q$ an isosbestic point \cite{isosb} appears  indicating the access to a stable regime of quantum fluctuations \cite{PhilMag} independent of the 
electronic environment variation.

In summary, the $C_P/T(T>T_m)$ dependencies in the paramagnetic region also indicate to which side of the QCP belongs the system: i) on the magnetic one, with a $C_P(T)/T\propto 1/T^p$ 
dependence and large values of $C_P/T|_{lim(T\to 0)}$, which are well distinguished from ii) the NFL with a $C_P/T\propto -log(T/T_0)$. iii) On the non magnetic limit are those behaving nearly temperature independent  $C_P/T$. 
In the following we will analyze the characteristic properties that identify each group.

\subsubsection{Systems approaching the QCP from the magnetic side}

\begin{figure}
\begin{center}
\includegraphics[width=20pc]{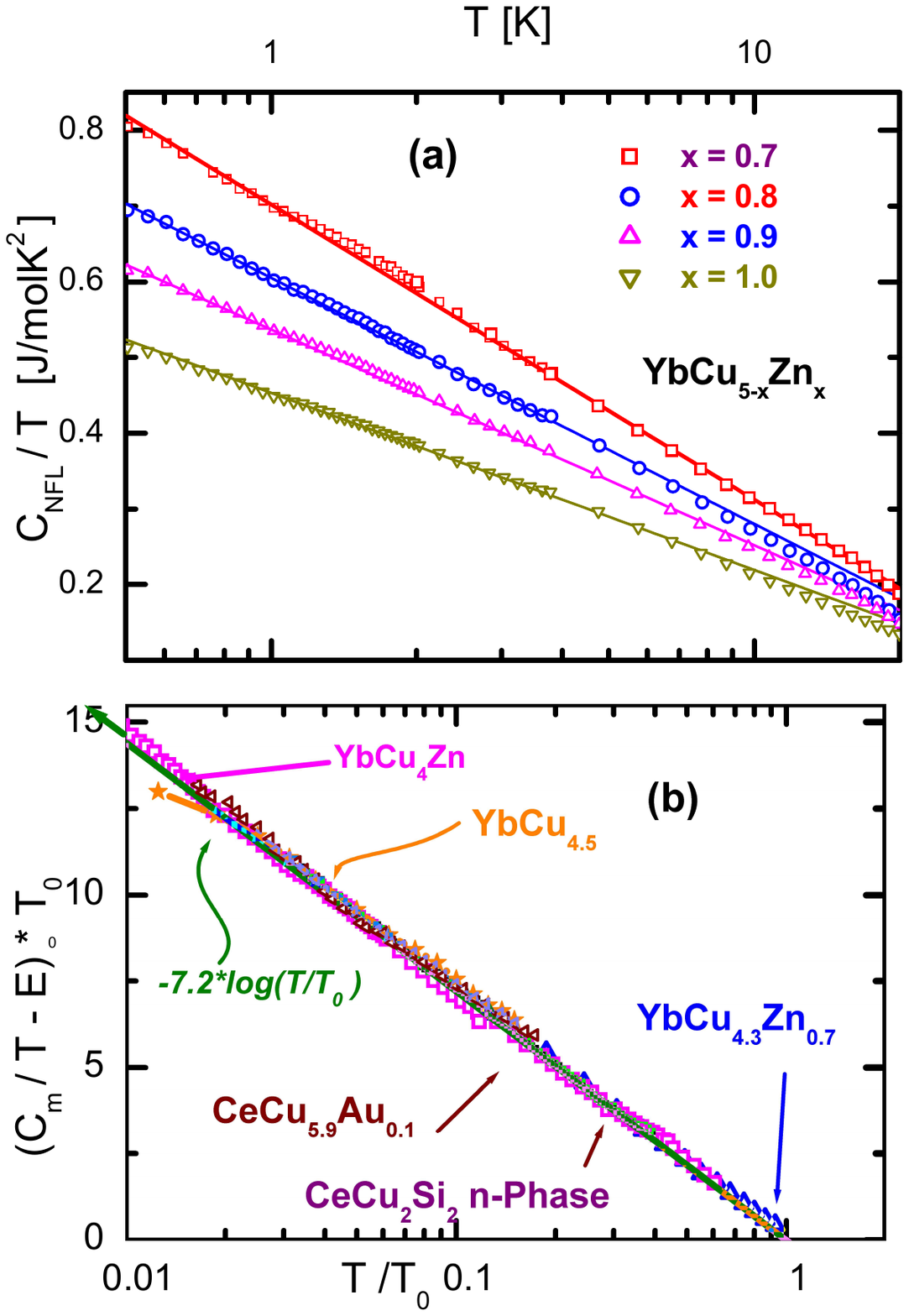}
\caption{(Color online) a) Logarithmic temperature dependence of YbCu$_{5-x}$Zn$_x$ alloys. b) Comparison of different NFL compounds within the normalized procedure \cite{NFL} including two Ce-referent compounds, after phonon subtraction (see text). \label{F4}}
\end{center}
\end{figure}

Hereafter we will take into account only the magnetic contribution $C_m(T)$ instead of the total 
specific heat $C_P(T)$ because for $T\leq 2,K$ the phonon contribution is irrelevant. 
The magnetic experimental limit is represented by YbNi$_4$Cd that shows a $C_m$ jump $\Delta C_m(T_{ord})=3/2R$, as a text book example for the mean field theory. This demonstrates that 
the full range from the 
magnetic limit is covered by this family of Yb-based compounds. The following compounds, YbCu$_4$Pd \cite{Pd} and YbCu$_4$Au \cite{YbCu4AuBau} show a sharp peak around 
$T=0.8$\,K. In Fig.~\ref{F2} one can see how the tail of $C_m(T>T_m)$ 
transforms into a power low temperature dependence $C_m \propto 1/T^p$ with $p=2.1$ for $x=1$, close to YbNi$_4$Mg that fit into that power law regime with $p=2$ \cite{YbNi4Mg}. This 
behavior coincides with that of YbCu$_{4.2}$Au$_{0.8}$, not shown for clarity. 

The '$p$' exponent of YbCu$_{5-x}$Au$_x$ alloys progressively decreases to down 
 $p=0.75$ for YbCu$_{4.6}$Au$_{0.4}$ \cite{Banda}. We remark that the studied  YbCu$_{5-x}
$Au$_{x}$ alloys with $p=0.9$ and 0.75 are exceptional cases of $p\leq 1$ exponents. 
It is worth to note that the $p\to 0$ extrapolation approaches a $log(T/T_0)$ function (to be discussed in the next subsection). Such power law dependence above $T_m$ was observed in 
system claimed to exhibit spin-liquid behavior \cite{SpinLiq, Materials, YbMgGa, Yb2Ti2O7}.

\begin{figure}
\begin{center}
\includegraphics[width=20pc]{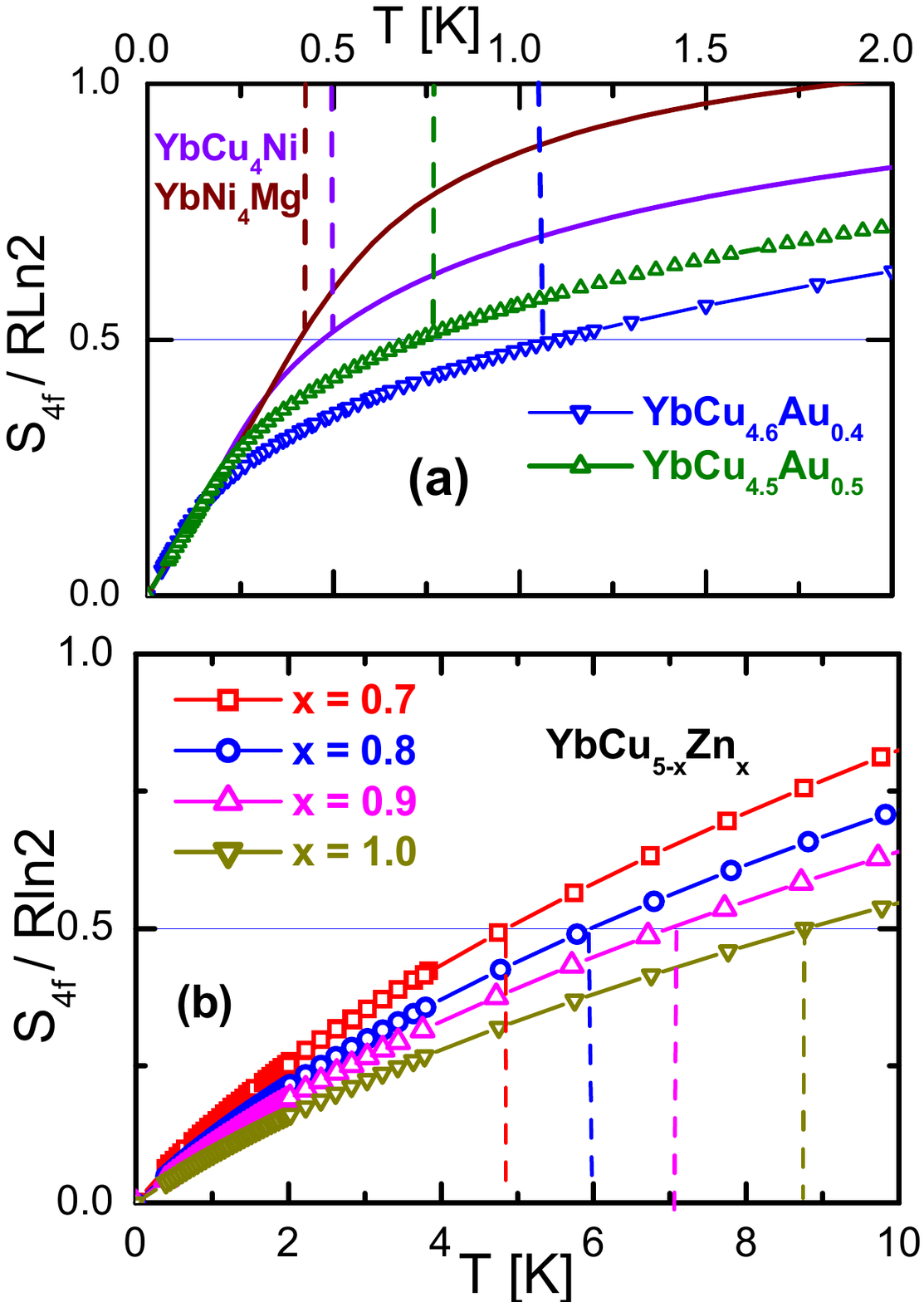}
\caption{(Color online) a) Evaluation of $T_K^{GS}$ of some magnetic alloys and compounds according to the $S_m(T_K)=2/3RLn(2)$ criterion.  b) Same procedure for the $T_K^{GS}$ 
determination in YbCu$_{5-x}$Zn$_x$ alloys. \label{F5}}
\end{center}
\end{figure}

Another relevant observation is that different curves with different $p$ exponents cross each other at an  {\it Isosbestic} point, see Fig.~\ref{F2} \cite{isosb} at $T=1.2$\,K and $C_p/T=1.05$\,J/
mol\,K$^2$. 
The fact that $C_P/T = \partial S/\partial T$ has the same value for different electronic configurations indicates that their behavior will not change as a function of the used {\it control 
parameter}: $C_m \equiv C_{\zeta}$ 
down to $T\to 0$ because their {\it internal energy} $U_m$ and {\it entropy} $S_m$ variation have found their minimum relation according to the $\partial U/ \partial S = T$ condition 
\cite{Abriata}. As a consequence, the temperature dependence of YbCu$_{5-x}$Au$_x$ with $x \leq 0.8$, together with YbCu$_4$Ni \cite{YbCu4Ni} and YbNi$_4$Mg, are governed by the same 
internal mechanism independently from the external control parameter variation r: i.e.  chemical doping in this case \cite{isosb, Abriata}.  

The evolution of the main parameters extracted from Fig.~\ref{F2} for these systems approaching the QCP are collected in Fig.~\ref{F3}. They are presented as a function of a control parameter $
\zeta$ that will be used 
in Section III and described in detail in the Supplementary Material. The accounted  compounds are the YbCu$_{5-x}$Au$_x$ alloys, including YbCu$_5$ as the $x\to 0$ limit, YbCu$_4$Ni and 
YbNi$_4$Cd. There is a clear coincidence between the rapid fall of  
$T_m(\zeta)$ and the exponent $p(\zeta)$ towards the QCP and the limit for the increase of $C_m/T|_{Lim T\to 0}(\zeta)$ at $\gamma_{max}$.

In Fig.~\ref{F3}, the decreasing transition temperature of YbCu$_{5-x}$Au$_x$ alloys 
from $T_m(x=1) =0.8$\,K to $T_Q(x=0.4) = 0.3$\,K \cite{YbCu5-xAux}, extrapolates to $T_Q \to 0$ at $x=0.32$ \cite{Banda}. 
Notably, the YbCu$_5$ compound, which does exist with the cF24-AuBe$_5$ structure, does not corresponds to this extrapolation because it forms under pressure,i.e. another external parameter 
than chemical doping. 
Coincidentally, the alloys YbCu$_{5-x}$Zn$_x$ form within the $0.7 \leq x \leq 1$ range of concentration \cite{ArXivZn}, and is a sort of 'mirror' system of YbCu$_{5-x}$Au$_x$ on the 
opposite side of the QCP. 

 It is worth noting that in the $T_m \geq T \ \geq T_Q$ region the $C_m/T|_{\lim{T \to 0}}$ limit starts growing from a quite low value up to the mentioned {\it plateau} with $\gamma_{max}$.  
 This means that by lowering 
 $T_m$ there is a continuous transference of degrees of freedom from the classical ordered phase to one dominated by quantum fluctuations. 
The occurrence of a {\it plateau}, with a coincident value of $C_m/T|_{\lim{T \to 0}}=\gamma_{max}\approx 6.7 \pm 0.4$\,J/mol K$^2$ for $T\leq T_m$. \cite{JLTP2018}, 
suggests a sort of limit for the $\partial S_{MO}/\partial T|_{\lim T \to 0} = C_m/T$ imposed by the Nernst´s principle. This striking feature, that was pointed out some years ago as an {\it Entropy 
Bottleneck} in many Ce-based compounds merits a deeper investigation because it was also observed in other non-Kramer GS like PrInAg$_2$ \cite{PrNiAg2} with the same $\gamma_{max}$ 
values, while Ce compounds show lower $\gamma_{max}(Ce)$ around 1\,J/molK$^2$. 
Within the quantum fluctuation scenario it can be expected that the linear $C_m|_{\lim{T \to 0}}(T)$ dependence  is originated in the quantum nature of this regime where the standard Boltzmann 
thermal distribution starts to compete with a tunneling effect between two minima of energy of the phases converging at the QCP. This scenario corresponds to a two-level model \cite{2level}, that  
dominates the $T\to 0$ physical behavior. 

The intensity of the Kondo interaction is evaluated according to the Desgranges-Schotte model \cite{Schotte} that defines the Kondo temperature as the temperature where $S_m(T)=2/3RLn2$, 
see Fig.~\ref{F5}. There one can appreciate how low are those values for the GS of the magnetic systems where $T_K^{GS}\approx 0.$\,K for  YbNi$_4$Cd and YbCu$_4$Mg; 0.96\,K for 
YbCu$_4$Ni; 1.6\,K for YbCu$_{4.5}$Au$_{0.5}$ and 2.3\,K for YbCu$_{4.6}$Au$_{0.4}$. The full change variation of $T_K^{GS}$ will be analyzed in the general phase diagram in Section III.

\subsubsection{Systems located at the non-fermi-liquid region}

The YbCu$_{5-x}$Zn$_x$ alloys included in Fig.~\ref{F4}a do not show any magnetic anomaly down to 50\,mK  even under magnetic field of $B=9$\,T \cite{Arxiv}, therefore they are placed on 
the {\it non-magnetic} side of the QCP. This positioning is not 
arbitrary because it is supported by a NFL: $C_m \propto log(T/T_0)$ dependence \cite{Steward}, where $T_0$ is a characteristic temperature related to the spin fluctuations energy, see  
Fig.~\ref{F4}a.

The NFL character of the $C_m(T)$ behavior of YbCu$_4$Zn can be verified using the general scaling criterion for NFL systems: $C_m / t = -D\, log(t) + E\,T_0$ \cite{NFL} shown in  Fig.~\ref{F4}b.
There, $t = T/T_0$ and $D = -7.2$\,J/molK. Since $D$ is a fixed factor of reference and $T_0$ is a  free parameter which provides an energy scale comparison between different systems. The term  $ET_0$ accounts for the conduction band 
contribution.  For YbCu$_4$Zn one obtains $T_0 = 33$\,K and $E= 90$\,mJ/mol\,K$^2$, that can be compared with the NFL reference systems: CeCu$_ {5.9}$Au$_{0.1}$ \cite{CeCu6Au} with $T_0 = 5.3$\,K and $E= 
53$\,mJ/Kmol, and the n-phase of CeCu$_2$Si$_2$ \cite{phaseN} with $T_0 = 14$\,K and $E=40$\,mJ/Kmol. Also for comparison, the Cu-rich alloy YbCu$_{4.3}$Zn$_{0.7}$  
\cite{Akbar} is included in the figure with $T_0 = 16$\,K and $E= 170$\,mJ/Kmol, indicating to be closer to the QCP than the stoichiometric compound YbCu$_4$Zn. In fact the 
standard $C_m(T)/T$ values measured above 2\,K are about 50\% larger for $x=0.7$ than those of $x=1$, see Fig.~\ref{F4}a, with $T_0$ twice smaller.
Also YbCu$_{4.5}$ \cite{Lilian4.5} is included in Fig.~\ref{F4}b, with a $T_0=40$\,K and  $E= 300$\,mJ/Kmol, though it is synthetized under pressure. 

The $C_m/T \propto log(T/T_0)$ dependence rarely appears in Yb-based compounds, nevertheless  the mentioned cases are not exceptions because with 19\% of Sc substitution in Yb$_{1-x}
$Sc$_x$Co$_2$Zn$_{20}$ it also shows a $log(T/T_0)$ dependence by approaching a QCP \cite{YbCo2Zn20}. 
The intensity of the Kondo interaction affecting the doublet-GS in this set of alloys is evaluated according the $S_m(T_K)=2/3RLn2$ criterion for the entropy as shown in Fig.~\ref{F5}b. In this 
case $T_K^{GS}$ ranges between 7\,K for $x=0.7$ and 14\,K for $x=1$, clearly larger than the values observed for the systems belonging to the magnetic side of the phase diagram. 

\subsubsection{Systems within the Valence Fluctuation regime}

The YbCu$_{5-x}$Ag$_x$ alloys included in Fig.~\ref{F6} are a sort of 'text book' example for Fermi-liquid behavior showing VF mechanism. 
As known Yb ions access to 
two electronic configurations: Yb$^{3+}$: [$Xe (6s^25d^1) 4f^{13}$] and Yb$^{2+}$: [$Xe (6s^15d^1) 4f^{14}$], being the former magnetic and the latter non-magnetic. After the localization of a $band$-
electron, which fills the $4f^{14}$ orbital, the atomic volume of $Yb^{2+}$ becomes significantly larger than that of $Yb^{3+}$ with $4f^{13}$ electrons. Therefore, the cell-volume of different Yb compounds is directly 
related to the variation of the electronic {\it band-4f} localization. This process occurs  
provided the system has a rich electronic-concentration from the ligand atoms and/or an increasing available volume \cite{available} for the Yb atom within the crystal structure position. The 
enhanced increase of atomic volume is referred to a sort of Vegard's law described in the Supplementary Material, see Fig.~\ref{F7}.

Besides the Kondo screening, once the $T_K^{GS}$ energy approaches that of the CEF splitting the VF phenomenon is observed like in Ce, Sm, Eu and Yb.
For this set of VF alloys, the $T_K^{GS}$ concept for a doublet-GS loses its meaning because $T_K$ overcomes the CEF splitting: $T_K>\Delta_{CEF}$ \cite{PhilMag}. Thus, the $T_K$ parameter 
fits into the original definition used for the D-L diagram \cite{Lavagna}. In that scenario, $T_K$ has to be evaluated at the temperature at which $S_m(T_K^{CEF})=2/3Rln(8)$, because the  
degeneracy of the total 
angular moment of Yb: $J=7/2$, is N=8. Since the available information about the specific heat of YbCu$_{5-x}$Ag$_x$ does not exceed $T\approx 15$\,K \cite{YbCuAgPRB} the $T_K$ 
extrapolation has a significant error bar. Such is not the case for YbCu$_5$ and YbCu$_4$Ag compounds because their respective entropy were determined quite precisely from $C_P/T$ 
measurements performed up to higher temperature with respective $T_K=17$\,K and 32\,K \cite{Hamdaoui}.

\section{Magnetic phase diagram}

\begin{figure*}
\begin{center}
\includegraphics[width=40pc]{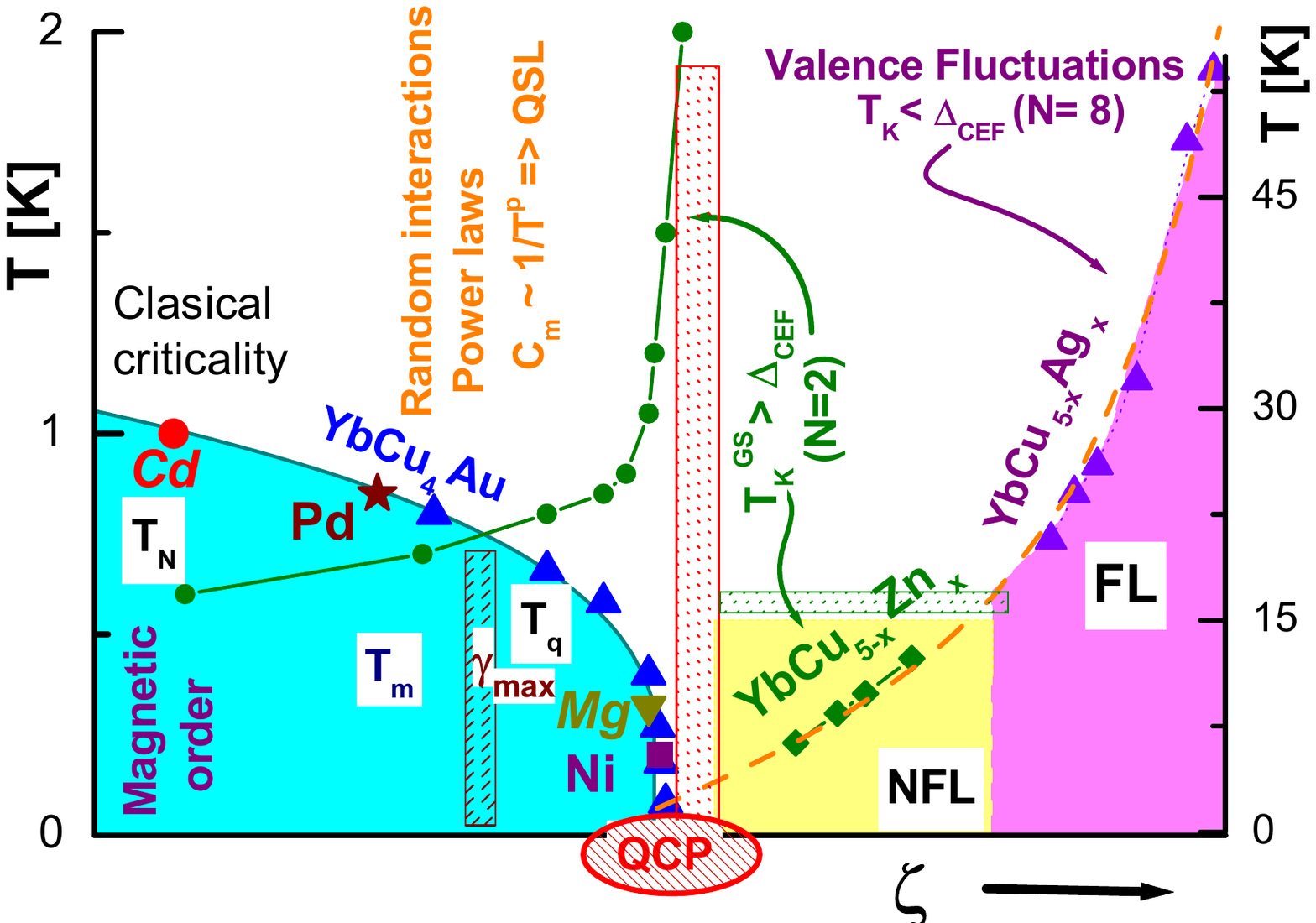}
\caption{(Color online) Schematic magnetic phase diagram in the vicinity of a QCP drawn according to the collected experimental information extracted from the YbM$_{5-x}$X$_x$ family as a 
function of {\it chemical doping} $\zeta$. Cyan region: magnetic interactions; yellow: non-Fermi-liquid NFL  behavior, and magenta: Fermi-liquid FL where $T_K^{GS} < \Delta_{CEF}$ splitting. 
Compounds with $M=$Ni are in italic. Note the different temperature scales between the magnetic and non-magnetic regions. 
 \label{F6}}
\end{center}
\end{figure*}

Addressing now to the construction of a magnetic phase diagram, one has to take int account that there is no one-to-one correspondence between the theoretic parameter $J_{ex}$ and any single 
control parameter. 
Measurements under hydrostatic pressure would fulfill this condition, but they use to cover a  limited range compared with the comprehensive study presented in this work. Neither external 
magnetic field does it because it mainly affects on the magnetic interaction symmetry 
flipping the magnetic moments projection without modifying their intensity. 
To compare the behavior of such a large number of compounds at hand within a broad range of behaviors we have used a control parameter $\zeta$ related to the {\it chemical potential}  
variation produced by chemical doping of the Yb-ligand atoms in YbM$_{5-x}$$_x$. The details about the $\zeta$ definition are included in the Supplementary Material section. 

A successful demonstration of the effect of the cell volume as a control parameter is found in the low temperature study on CeCu$_{6-x}$Au$_x$ ($0.3 \geq x \geq 0$) alloys \cite{CeCu6Au} that 
shows how an alloy can be driven from a magnetic to a non-magnetic GS when major Au atoms are substituted by smaller isoelectronic Cu within the NFL region included in Fig.~\ref{F4}b. Similar 
effect is observed under applied pressure on the alloy CeCu$_{5.9}$Au$_{0.1}$, tuned close to the critical concentration. In comparison, in the same alloy the magnetic field induces a polarization of 
the magnetic structure without undergoing any clear NFL region. 

Based on all the experimental information collected on this rich family of Yb-based compounds around the QCP region, one can build a realistic magnetic phase diagram like the presented in Fig.~\ref{F6}. It condenses the most relevant features observed on both sides of the critical 
point and reveals a number of new properties not previously reported. 

\subsection{Description of the magnetic phase diagram}

 This phase diagram contains two well differentiated regions: the so called {\it magnetic} (on the left) and the {\it non-magnetic} (on the right) with different temperature scales in respective axes. 
 The  magnetic group includes  the Yb systems with relevant but decreasing magnetic interactions (cyan region) that covers from long range magnetic order, with a $C_m(T_N)=3/2R$ mean field 
 jump at $T_N=1$\,K in YbNi$_4$Cd \cite{YbNi4Cd}, to a cusp at $T_m$ dominated by magnetic fluctuations. Decreasing the temperature that cusp transforms into a shoulder at $T=T_Q$ between 
 a {\it plateau} in $C_m/T(T<T_Q)$ and a {\it spin-liquid} power law above it. The change of nomenclature from $T_{ord}$ to $T_m$ and then to $T_Q$ is introduced to indicate clear changes 
 of regime reflected in the magnetic phase boundary by approaching the QCP.
The $T_Q(\zeta)$ curve drops faster once $T_K^{GS}$ starts growing from very low values: $\approx 0.6$\,K in YbNi$_4$Mg, YbCu$_4$Ni \cite{YbNi4Mg} and YbCu$_{4.2}$Au$_{0.8}$.

Note that such incipient growing of $T_K^{GS}$ in a narrow region on the left side of the critical $\zeta$ value  coexists with the so-called {\it plateau} region, where $C_P/T|_{lim T\to 0}
=\gamma_{max}$. The question arises about the physical conditions producing such a saturation of $\gamma(\zeta)$ at $\gamma_{max}=6.7\pm 0.4$\,J/molK$^2$ and about its universality. The 
{\it plateau} character seems to be related with the formation of a continuous spectrum of excitations as it is observed in a two levels scenario with random energy distribution of energy minima \cite{2level}. 
Such a large density of excitations can be related to quantum fluctuations acting on a multiple energy gaps distribution where the system may tunnel between two minima of energy. 
It is worth noting that the same large value of $\gamma_{max}$ is also observed in a Pr compound \cite{PrNiAg2, PhilMag} indicating a sort of universality 
of this value. 

 On the non-magnetic side, two well defined regions can be distinguished in accordance to the GS degeneracy. Close to the critical point the scenario is dominated by a NFL behavior.
The energy scales obtained from magnetic, thermal and transport measurements on the exemplary alloys YbCu$_{5-x}$Zn$_x$ indicate that the characteristic energy of such  
fluctuations is $T_0 \approx 8$\,K \cite{YbCu4Zn}. 
There, the scenario is dominated by competing magnetic and quantum fluctuations. For moderated values of $T_K^{GS}$, i.e. much lower that $\Delta_{CEF}$ splitting, the 
N=2 degeneracy of the GS is kept and the exemplary system YbCu$_{5-x}$Zn$_x$ behaves as a NFL as shown in  Fig.~\ref{F4}a. 
The compound YbCu$_{4.5}$, which also shows a NFL logarithmic $T$  dependence \cite{Lilian4.5}, fits properly into the universal curve for NFL systems presented in Fig.~\ref{F4}b. 

As mentioned before, $T_K^{GS}$ is evaluated as the temperature at which $S_{GS}(T_K^{GS}) = 2/3 R\ln(2)$ \cite{Schotte} for a doublet GS.
However, for systems behaving as VF all CEF levels (N=8 for $J=7/2$) are involved in the GS properties because $T_K^{CEF}>\Delta_{CEF}$ \cite{PhilMag}. Therefore it has to be evaluated as  
$S_{CEF}(T_K^{CEF}) = 2/3 R\ln(8)$.
The exemplary case for this group is the YbCu$_{5-x}$Ag$_x$ family of alloys. Interestingly, the $T_K^{CEF}$ curve can be seen as an extension of the $T_K^{SG}$ at low energy. The YbCu$_5$ 
compound in its cubic structure fits into this curve among the YbCu$_{5-x}$Ag$_x$ alloys according to X-Ray-Absorption measurements that reveals an intermediate valance of 2.73 for Yb 
\cite{Hamdaoui}.

\section{Conclusions}

We have profited of the large number of Yb-based compounds and alloys showing the same cubic AuBe$_5$ type structure to study the physical evolution of magnetic compounds undergoing a QCP 
region. The 
broad range of magnetic behavior covered by this family of compounds offers a unique possibility because the previously investigated systems only form in a limited range of the applied control parameters. 
A significant number of unexpected features are revealed by this study around the critical region investigated down to the $T<1$\,K range of temperature.

Within the D-L model, we have recognized the difference between the magnetic coupling parameter $J_{ex}\propto T_N$ amid well defined magnetic moments, and the {\it 4f-band}  
electron state interaction $J_K$ which weakens the magnetic moments intensity. Therefore, for the systems expected to access to a QCP following theoretic hypothesis of a phase transition reaches 
zero temperature it is necessary that $T_N\approx T_K$. This realistic 
condition is only fulfilled by a magnetically robust doublet-GS state originated in the CEF effect, accounting that its Kondo energy $T_K^{GS} << \Delta_{CEF}$ splitting. 

To drive the transition to zero by weakening the exchange interactions between magnetic moments, without a significant weakening of their intensities requires the action of another 
mechanism, like e.g magnetic or topological frustration. The thermal energy range where the vicinity to a QCP starts to affect the physical behavior corresponds to the $T_m$ region, where the 
increase of $C_m/T|_{\lim 
\to 0}(\zeta)$ up to $\gamma_{max}$ reflects the progressive transference of degrees of freedom from classical magnetic interactions to quantum fluctuations. This occurs  between  $ \approx 1$\,K 
and $\approx 0.4$\,K. Below that temperature quantum effects governs the scenario conditioned by the third law of thermodynamic constraints on the entropy evolution at $T\to 0$.
Coincidentally, in that region a power law $C_m/T(T>T_m)$ tail is observed, with a decreasing exponent of $1/T^p$ from $\approx 2$ to 0.75 together and with an {\it isosbestic} point at $T\approx1$\,K. 
The universal $\gamma_{max}$ value can be related to the quantum tunneling effect within a 
random distribution of energy gaps between two levels. 

At the QCP, a discontinuity in $C_m/T|_{\lim \to 0}(\zeta)$ occurs, followed by a NFL temperature dependence with significantly lower $\gamma(\zeta)$.
Then, it progressively grows again where $T_K^{GS}$ value approaches the CEF splitting energy. This behavior was already highlighted in Ce alloys covering similar range from magnetic (doublet-
GS) to valence fluctuation (N=8 GS) evolution \cite{Physica95}. 
At the onset of the FL, the entropy definition of the doubet-GS Kondo temperature: $S_m(T_K^{GS})=2/3\ln(2)$, loses its meaning because all the Hund´s rule levels for $J=7/2$ have to be taken into account, thus $S_m(T_K)=2/3\ln(8)$ \cite{PhilMag, JLTP2018}. 

A comparison between the $C_m(T)$ behavior analyzed in this work and the corresponding magnetic susceptibility and electrical resistivity within the same range of temperature is highly 
desirable.

\newpage 

\section*{Supplementary Material}

\begin{figure}
\begin{center}
\includegraphics[width=19pc]{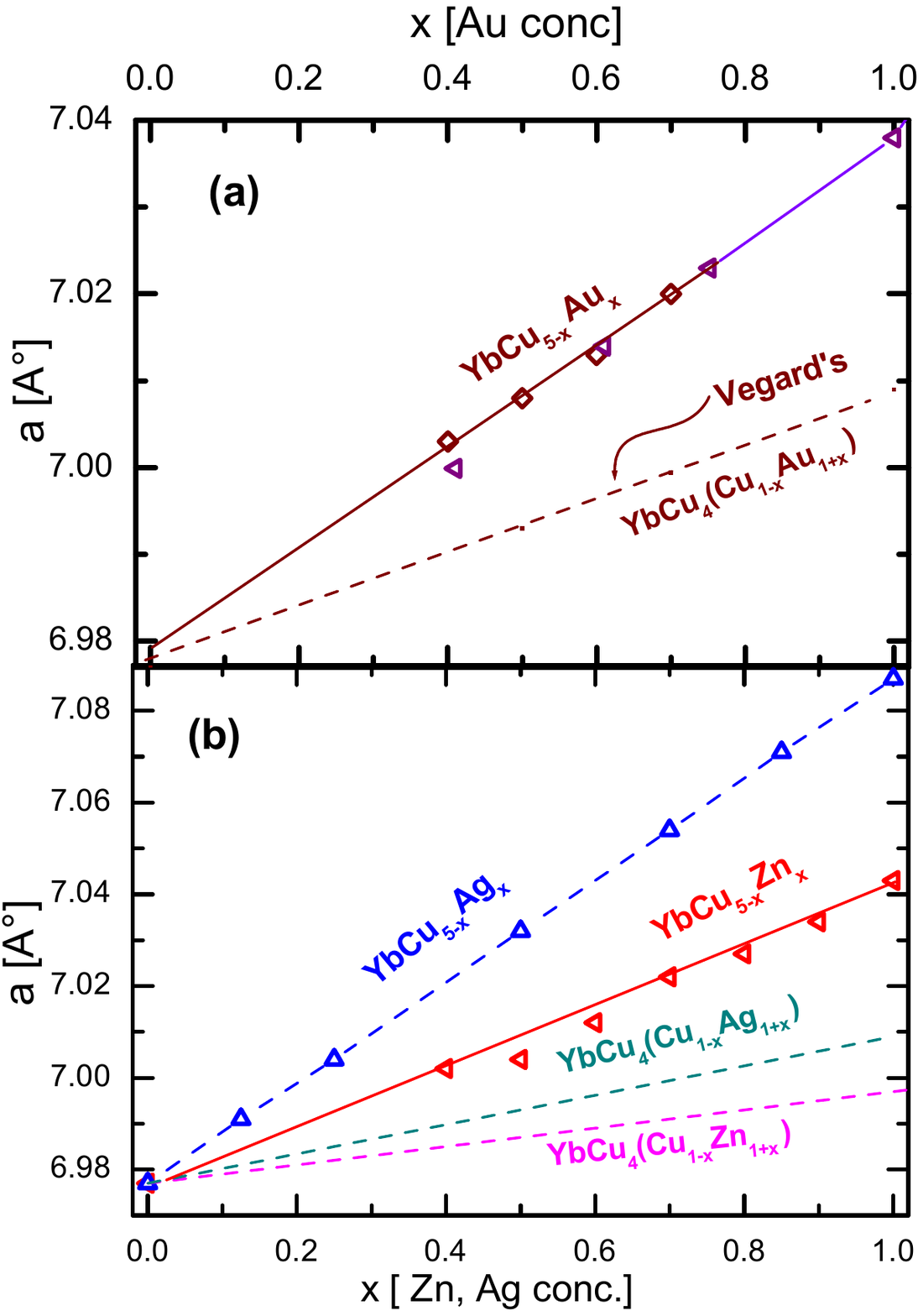}
\caption{(Color online) a) Comparison of the lattice parameter variation of YbCu$_{5-x}$Au$_x$  \cite{YbCu4Au} with an hypothetical Vegard's law reference, where one (of five) Cu atoms is 
progressively substituted by one Au. The $x=0$ origin in the figure is taken from YbCu$_5$ \cite{YbCu5}. b) The same for YbCu$_{5-x}$Zn$_x$  \cite{Akbar} and YbCu$_{5-x}$Ag$_x$ 
\cite{Yoshi97} alloys respective Vegard's references.
\label{F7}}
\end{center}
\end{figure}

\subsection{Sturcture, lattice prameters, Vegard's law} 

Respective lattice parameters of the studied compounds are: a(YbCu$_4$Ni)= 6.943\,\AA\     
\cite{YbCu4Ni}, a(YbCu$_4$Pd) =7.039\,\AA \cite{Pd}, a(YbCu$_4$Au)=7.046\,\AA  \cite{YbCu4AuBau,YbCu5-xAux} and alloys \cite{YbCu4Au}, a(YbCu$_4$ Zn)=7.0399\,\AA\ and 
alloys, \cite{Akbar}, 
a(YbNi$_4$Mg)= 7.032\,\AA \cite{YbNi4MgLatPar}, a(YbNi$_4$ Cd)=6.975\,\AA  \cite{YbNi4Cd}, a(YbCu$_4$Ag)=7.082\,\AA \cite{Besnus,Hamdaoui} and alloys \cite{Yoshi97}, and a(cubic YbCu$_5$)= 6.969\,\AA \cite{Hamdaoui}. 

\begin{figure}
\begin{center}
\includegraphics[width=16pc]{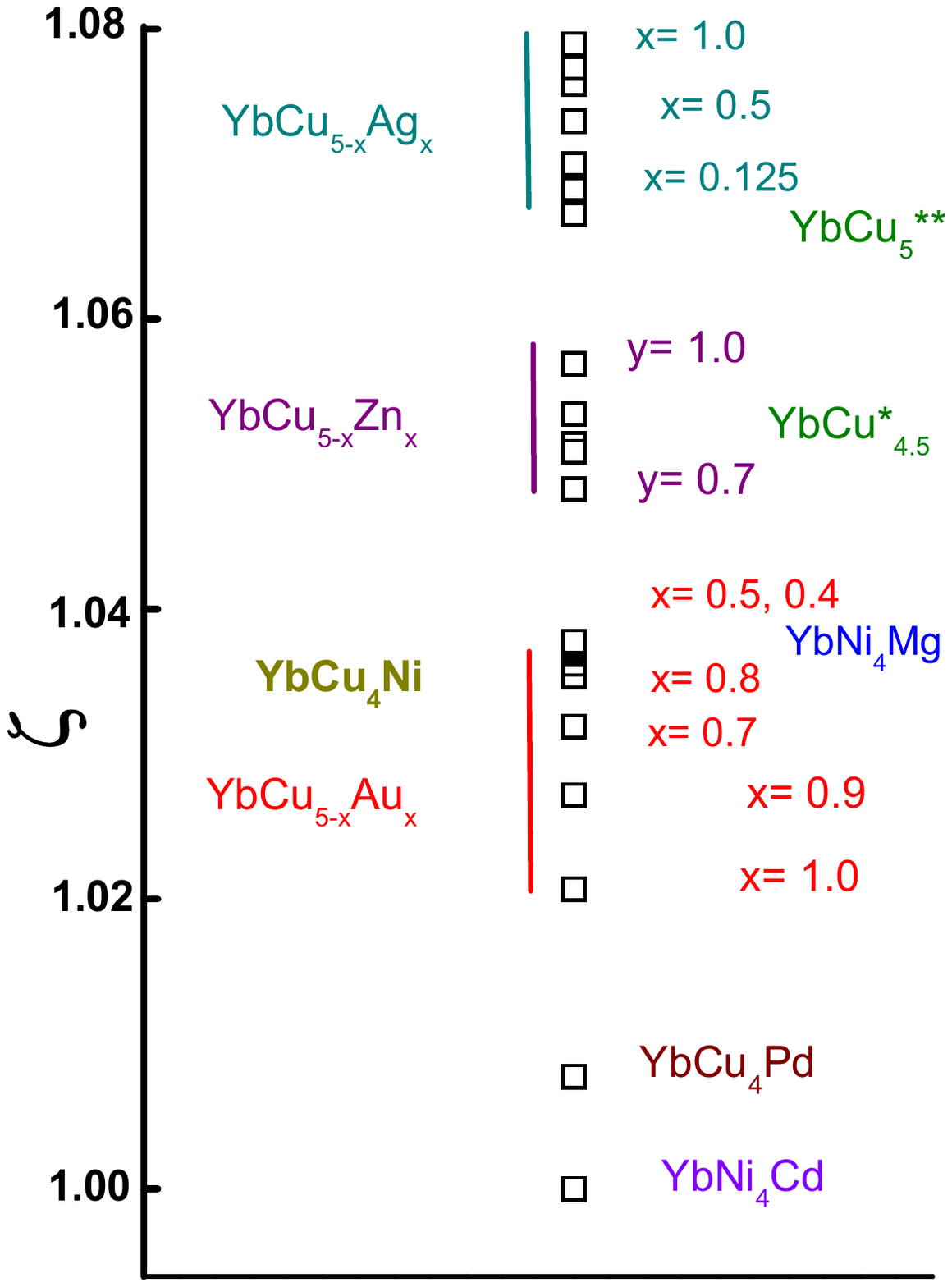}
\caption{Scale of $\zeta$ values, see the text, for different compounds and alloys }.
\label{F8}
\end{center}
\end{figure}

In Fig.~\ref{F7} we compare the variation of the lattice parameters of three YbCu$_{5-x}$X$_x$ alloys ($X$ = Au,  Zn or Ag) with hypothetical reference alloys that follow the Vegard's law 
between stoichiometric limits: YbCu$_5$ 
and respective YbCu$_4X$. Those computed lattice parameters are computed progressively substituting one (of five) Cu atoms by one corresponding $M$ atom, like in formula: YbCu$_{5-x}
X_x$. We note that the $X$=Ag alloys 
show the largest increase of lattice parameter, which clearly exceeds that of the corresponding Vegard's law. This is due to the fact that there is an extra progressive increase of the Yb volume 
from the smaller magnetic Yb$^{3+}$ to the larger non-magnetic Yb$^{2+}$ configuration.

\subsection{Control parameter} 

To design a phase diagram encompassing such large spectrum of magnetic behaviors displayed by the Yb$M_{5-x}X_x$ family, we have used a sort of {\it chemical 
potential} as control parameter driven by chemical doping, that at $T\to 0$ coincides with the Fermi energy. They depend on  the Wigner-Size concept, which involve the cell volume $V$ and the 
electronic configurations of Yb and its ligand atoms. In Yb-based systems, the chemical potential is relevant because it can access to two different electronic configurations discussed in Subsection II.2, 
being the smaller Yb$^{3+}$ magnetic, and the larger Yb$^{2+}$ non-magnetic. This is related to the localization of a {\it band-electron} into the $4f$ orbitals fulfilling the $4f^{14}$ shell. 
Therefore, according with the atomic 'available volume' \cite{available}: larger size and higher Fermi-energy favors such localization. 

In our case, the first approach to define a common control parameter $\zeta$ \cite{zeta} 
is done by sorting all components of this Yb-based family according to their respective cell volumes $V$ within the same cubic structure. For practicality, the cell volumes are normalized to the volume of the most 
magnetic compound: $V_0$ of YbNi$_4$Cd: i.e. $ \Omega = V/V_0$. 
Although that volumetric parameter works properly as a first approach to sort alloys with the same Yb-ligands, to compare systems with different $M$ or $X$ atoms it also requires to take into  
account the differences between the electronic concentrations of the involved atoms.

Thus, the heuristic criterion applied to converge into a unique control parameter: $\zeta = (1+b) \Omega$, where '$b$' introduces a correction factor related to the difference between 
the electronic concentration of different Yb-ligands in order to obtain a continuous $T_{ord}(\zeta)$ and $T_K(\zeta)$ variation, as it was presented in the magnetic phase diagram of Fig.~\ref{F6}. 
The correction represented by '$b$' ranges between $0\leq b \leq 5.4\%$, being the maximum for  YbCu$_4$Ni. 
In Fig.~\ref{F8} we present all studied systems sorted according their respective $\zeta$ paramenter values.

\end{document}